\documentclass[%
 aip,
 pop,
 amsmath,amssymb,
 reprint,%
]{revtex4-1}
\usepackage{graphicx}% Include figure files
\usepackage{dcolumn}% Align table columns on decimal point
\usepackage{bm}% bold math
\usepackage{overpic}
\usepackage{tikz}
\usepackage{color}

\usepackage[utf8]{inputenc}
\usepackage[T1]{fontenc}
\usepackage{etoolbox}

\newcommand{\BB}{\mathbf{B}}

\newcommand{\xxc}{\mathbf{x}_{\rm c}}
\newcommand{\xx}{\mathbf{x}}
\newcommand{\nn}{\mathbf{n}}
\newcommand{\JJ}{\mathsf{\mathsf{J}}}
\newcommand{\SG}{\mathsf{\mathsf{\Sigma}}}
\newcommand{\UU}{\mathsf{\mathsf{U}}}
\newcommand{\VV}{\mathsf{\mathsf{V}}^*}
\newcommand{\RR}{\mathcal{R}}

\makeatletter
\def\@email#1#2{%
 \endgroup
 \patchcmd{\titleblock@produce}
  {\frontmatter@RRAPformat}
  {\frontmatter@RRAPformat{\produce@RRAP{*#1\href{mailto:#2}{#2}}}\frontmatter@RRAPformat}
  {}{}
}%
\makeatother
\begin{document}

\preprint{AIP/123-QED}

\title{Efficient single-stage optimization of islands in finite-$\beta$ stellarator equilibria}

\author{C. B. Smiet}
\email{Christoper.smiet@epfl.ch}
\author{J. Loizu}%
\author{E. Balkovic}%
\affiliation{ 
Ecole Polytechnique Federale de Lausanne, Lausanne, Switzerland
}%
\author{A. Baillod}
\affiliation{%
Department of Applied Physics and Applied Mathematics, Columbia University, New York, New York 10027, USA
}%

\date{\today}% It is always \today, today,

\begin{abstract}
We present the first single-stage optimization of islands in finite-$\beta$ stellarator equilibria. Stellarator optimization is traditionally performed as a two-stage process; in the first stage, an optimal equilibrium is calculated which balances a set of competing constraints, and in the second stage a set of coils is found that supports said equilibrium. Stage one is generally performed using a representation for the equilibrium that assumes nestedness of flux surfaces, even though this is not warranted and occasionally undesired. The second stage optimization of coils is never perfect, and the mismatch leads to worse performing equilibria, and further deteriorates if additional constraints such as force minimization, coil torsion or port access are included. The higher fidelity of single-stage optimization is especially important for the optimization of islands as these are incredibly sensitive to changes in the field.  In this paper we demonstrate an optimization scheme capable of optimizing islands in finite $\beta$ stellarator equilibria directly from coils. We furthermore develop and demonstrate a method to reduce the dimensionality of the single-stage optimization problem to that of the first stage in the two-stage approach. 
\end{abstract}

\maketitle

One of the challenges in stellarator design is the vastness of the design space; encompassing families of entirely different magnetic configurations including but not limited to (quasi)-isodynamic configurations~\cite{goodman2023constructing}, quasisymmetric configurations~\cite{rodriguez2022quasisymmetry,rodriguez2022measures}, and isodrastic configurations~\cite{burby2023isodrastic}. 
Designing a stellarator equilibrium can be conceptualized as an optimization in this infinite-dimensional space of possibilities, which is a challenge in its own right, but has proven successful in experiment~\cite{beidler2021demonstration}, and yielded many interesting configurations~\cite{goodman2023constructing, landreman2022magnetic}. 

The magnetic field supporting the plasma equilibrium needs to be generated by a set of coils external to it, which in a two-stage process is designed a posteriori. 
This is done by matching the coil solution to the plasma solution on the plasma boundary, by minimizing the squared total magnetic flux through this boundary.
The field generated by a finite number of discrete coils can closely approximate the required field, but the match will never be exact. 
Reduction of the total coil length exponentially increases the error in this match due to coil ripple, and destroys the perfect confinement characteristics of the optimized equilibrium~\cite{wechsung2022precise}. 
Construction costs on the other hand increase with the quantity of conductor needed, and are reduced by having fewer total coils of shorter length.

The requirements of a reactor place constraints on the coils that can potentially degrade equilibrium performance but can be mitigated through a single-stage approach.
Neutral beams require locations with increased spacing between the coils, view ports and feedtroughs must be placed where they are needed, and device maintenance considerations set aditional constraints on the coil geometry~\cite{brown2015engineering}.
A better reactor can be designed if the coils are directly included in the optimization, such that these trade-offs can be weighed, and so that the unavoidable coil ripple is distributed in a way to minimize the deterioration of performance.

Single stage optimization, where the coils are included during the equilibrium optimization, can be achieved in several different ways, comprehensively presented in~\cite{henneberg2021combined}. 
The most precise, but also costly method is the direct-from-coil optimization where the coils are directly varied and the equilibrium supported by those coils is optimized. 
Recently Jorge~\emph{et al.} presented optimizations where a finite-$\beta$ fixed-boundary equilibrium was optimized in parallel with the coils, achieving better performing stellarator equilibria than equivalent two-stage optimizations~\cite{jorge2023single}. 
Another single-stage optimimzation method was reported by~\citet{giuliani2022direct} which utilizes adjoint methods to rapidly optimize vacuum fields for quasisymmetry. 
The methods' efficiency is attested by the 300,000 (and counting) coil configurations made available in the open-source QUASR database~\cite{giuliani2024direct}. 

When optimizing a stellarator equilibrium it is important to consider magnetic islands, which can reduce the confinement in the core. 
Many equilibrium solvers, including the widely used VMEC~\cite{hirshman1983steepest}, are unable to represent magnetic islands. 
This can be overcome by coupling a code capable of resolving islands and chaotic fields to the optimization as demonstrated by~\citet{landreman2021stellarator} for vacuum configurations ($\beta=0$). 
Islands are extremely sensitive to small changes in the magnetic field as was demonstrated in W7-X, where a field perturbation as small as $1:10^{-5}$ caused a discernible change in the island structure~\cite{pedersen2016confirmation}.

Whereas magnetic islands are detrimental to confinement in the core, they can be essential in the plasma edge, where a reactor furthermore requires a controlled method of handling the heat exhausted from the plasma. 
This can be achieved by an island divertor where a large island chain is intentionally generated, or a nonresonant divertor~\cite{bader2017hsx} where the last closed flux surface transitions into a chaotic field. 
To determine the points where the plasma interfaces with the wall, and hence the diveror geometry, the equilibrium representation needs to represent more than nested surfaces. 
Furthermore, ensuring that the coils do not interfere with the divertor can only be accomplished by a single-stage optimization that is capable of resolving magnetic islands and magnetic chaos. 

In this paper we present a first demonstration of a single-stage optimization capable of addressing all the issues above: Optimizing a free-boundary finite-$\beta$ equilibrium directly from and self-consistent with a set of coils, and targeting properties of the islands in the optimization.
We do this by removing the islands from around a $\langle\beta\rangle=1.49\%$ rotating ellipse equilibrium. 
We build upon our recent work~\cite{baillod2022stellarator, baillod2023equilibrium, baillod2023thesis} where magnetic islands in a finite-$\beta$ free-boundary equibrium were optimized without consideration of coils. 
In this paper we extend this to direct-from-coil optimization of islands in stellarator equilibria. 
We further demonstrate a dimensionality-reduction scheme that reduces the complexity and cost of the fully single-stage optimization to that of the direct free-boundary optimization.

\section{MR$\textsc{x}$MHD Equilibria coupled to coils}\label{sec:coupling}

In this work we use the Multi-region Relaxed MagnetoHydroDynamics (MRxMHD)~\cite{bruno1996existence} equilibrium formulation, solved by the Stepped Pressure Equilibrium Code (SPEC)~\cite{hudson2011non} to describe stellarator equilibria. 
An MRxMHD equilibrium consists of a number of discrete and nested ideal interfaces which sustain a finite pressure jump, and a force-free field in between. 
Unlike formulations such as VMEC~\cite{hirshman1983steepest}, that enforce continuously nested flux surfaces in their representations MRxMHD allows for solutions that contain magnetic islands and magnetic field line chaos whilst still constraining pressure and current profiles. 
This makes SPEC a powerful tool to predict equilibrium $\beta$ limits~\cite{baillod2023equilibrium} and study linear and nonlinear, ideal and resistive instabilities~\cite{loizu2023nonlinear, loizu2020direct, kumar2021computation}, and helical states in reversed field configurations~\cite{dennis2014modeling, liu2024effects}.

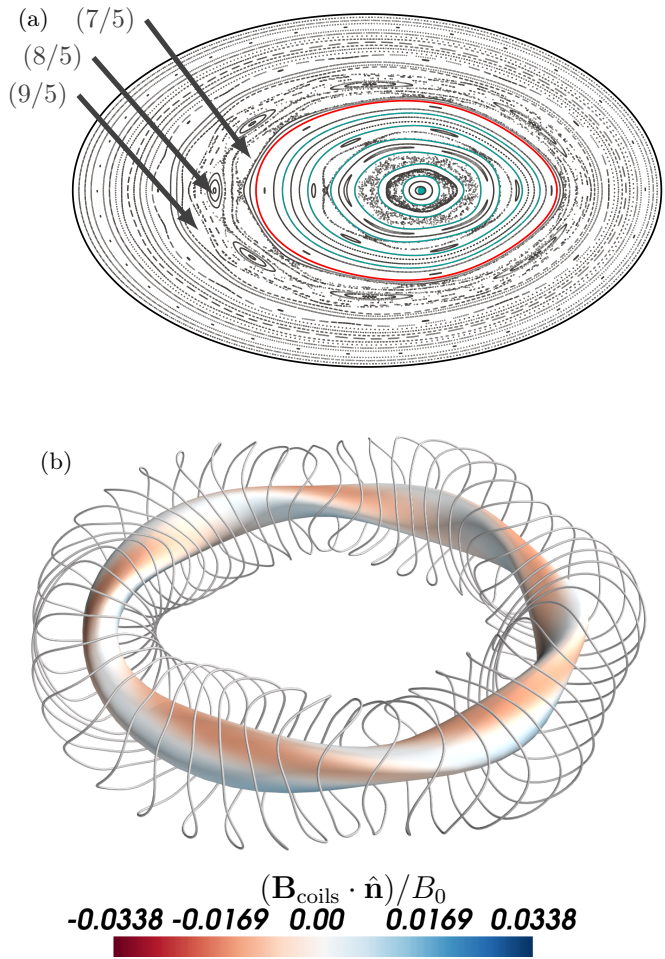
\begin{figure}
\begin{tikzpicture}
\node [anchor=south west] (image) at (0,0) {\includegraphics[width=\linewidth]{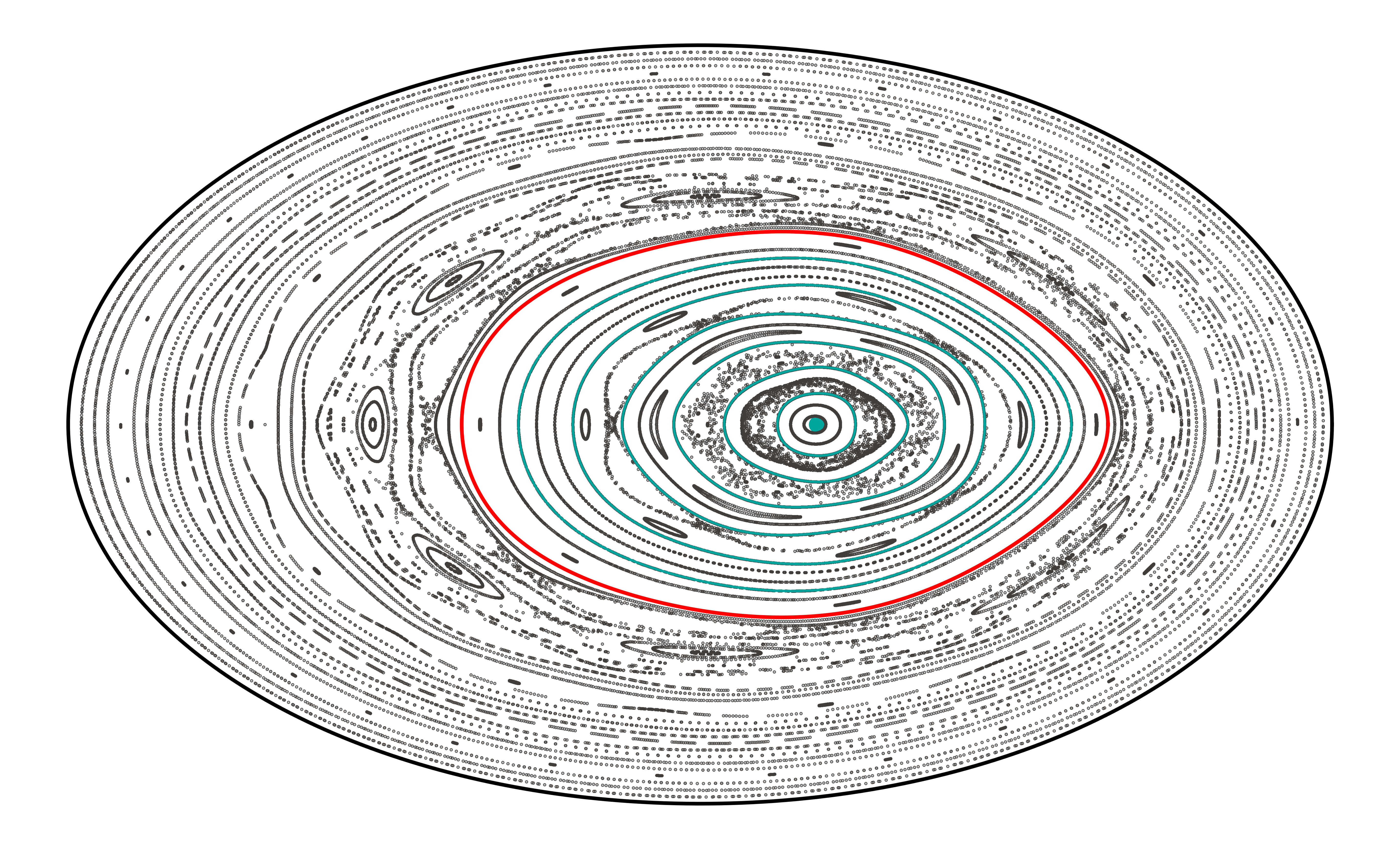}};
\coordinate (A) at (2.90,3.15); %\coordinate (A) at (7.95,10.3); 1.1; 1.45
\coordinate (Ap) at (1.5,5.0);
\draw[-latex,darkgray, line width=2pt] 
    node [left, font=\normalsize] at (Ap) {$(7/5)$} 
    (Ap) -- (A);
\coordinate (B) at (2.4,2.67);
\coordinate (Bp) at (0.8,4.50);
\draw[-latex,darkgray, line width=2pt] 
    node [left, font=\normalsize] at (Bp) {$(8/5)$} 
    (Bp) -- (B);
\coordinate (C) at (2.2,2.2);
\coordinate (Cp) at (0.6,4.0);
\draw[-latex,darkgray, line width=2pt] 
    node [left, font=\normalsize] at (Cp) {$(9/5)$} 
    (Cp) -- (C);
  \node[] at (0, 5.0) {(a)};
% \begin{scope}[x={(image.south east)}, y={(image.north west)}]
%  \coordinate (target) at (0.5, 0.7)
%  \draw [->, red, text width=2cm] ([shift((0,-0.2cm))!island a]target) -- target;
% \end{scope}
\end{tikzpicture}
  \begin{overpic}[width=\linewidth, percent]{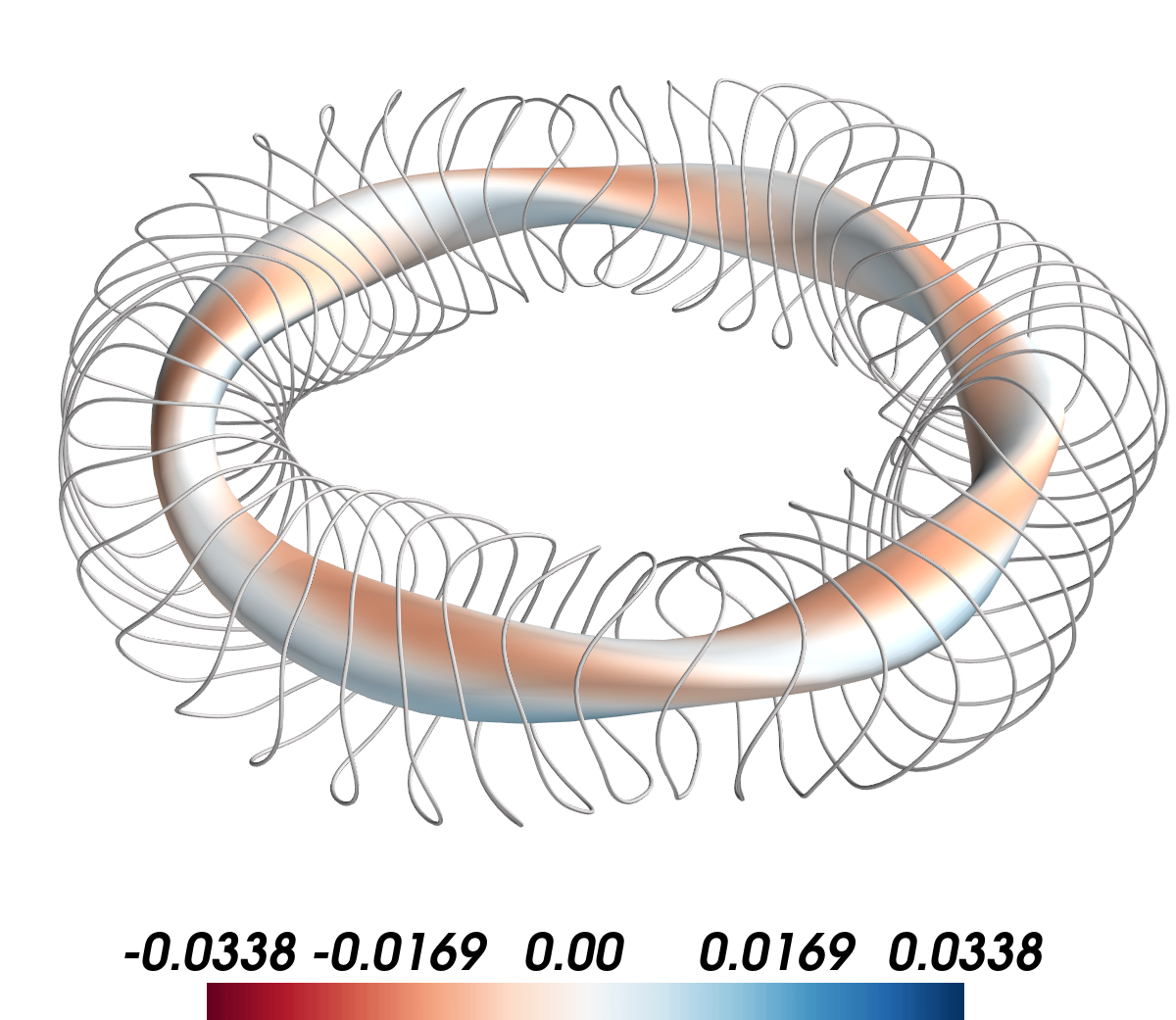}
    \put(5,75){ (b)}
    \put(40,9){\large$(\BB_{\rm coils}\cdot\hat{\mathbf{n}})/B_0$}
  \end{overpic}
  \caption{Initial unoptimized configuration and coils supporting this equilibrium. a: Poincar\'e section of the field, with the $(m/n)=(7/5)$, $(8/5)$ and $(9/5)$ islands labeled. The SPEC interfaces are from outside in: Computational boundary (black), plasma boundary (red), 6 internal interfaces (teal) and the magnetic axis (teal dot).  b: coils consistent with the equilibrium in (a), and the normal component of the coil magnetic field on the computational boundary. The color scale is identical to figures~\ref{fig:bruteopt}-\ref{fig:twostage} and is normalized to $B_0$, the magnetic field strength on the magnetic axis.}
  \label{fig:initeq}
\end{figure}

We take as starting point the same equilibrium presented in~\citet{baillod2022stellarator}, namely a rotating-ellipse field in which the pressure has been increased to $\langle\beta\rangle=1.49\%$, shown in figure~\ref{fig:initeq} (a). 
This equilibrium has 5 field periods, is stellarator-symmetric~\cite{dewar1998stellarator}, has 7 pressure interfaces, and is contained in a computational boundary with major radius 10m, minor radius 1m and an elongation of 2.5.
For a full description we refer to~\cite{baillod2022stellarator, baillod2023equilibrium}. 

In free-boundary mode, SPEC requires as an input the normal component of an externally produced magnetic field $(\BB_{\rm ext}\cdot\mathbf{n})$ on a computational boundary. 
SPEC uses a spectral representation, here with a poloidal resolution $M=6$ and toroidal resoltuion $N=8$. 
This means all doubly-periodic odd functions (and in particular, the normal component of the field on the boundary  $(\BB_{\rm ext}\cdot\mathbf{n})(\theta, \phi)$~) are represented using a double Fourier sine series of the form: 
\begin{multline}\label{eq:FourierDef}
  (\BB_{\rm ext}\cdot\mathbf{n})(\theta, \phi) = \sum_{n=1}^N V_{0,n}\sin(-nN_P\phi) + \\ \sum_{m=1}^{M}\sum_{n=-N}^N V_{m,n}\sin(m\theta-nN_P\phi)
\end{multline}
(even functions are described by the analogous cosine series). 
Here $\phi$ is a toroidal angle, $\theta$ a poloidal angle, and $N_P$ the number of field periods. 

In \citet{baillod2022stellarator} an optimzization was performed using as degrees-of-freedom the representation of the external field, which with the resolution $M=6$, $N=8$ constitutes the 110 Fourier coefficients of equation~\ref{eq:FourierDef}.
In a single-stage optimization, these Fourier components are not degrees-of-freedom, but they are determined by the coils. 
The degrees-of-freedom then become the parameters determining the geometry of the coils, from which the normal field to the boundary is calculated using the Biot-Savart law.

%We choose to represent our stellarator using 8 volumes and 8 interfaces, and is contained within the computational boundary consisting of a rotating ellipse with major radius 10$m$, minor radius 1$m$ and an elongation of 2.5.
%The equilibrium has 5 field periods and the field is stellarator-symmetric~\cite{dewar1998stellarator}. 
%The toroidal fluxes in each volume scale quadratically such that the interfaces are approximately equidistant. 
%We assume no externally driven currents by setting the volume currents $I_\mathcal{V} = 0$, and assume the plasma generates a bootstrap-like current on the pressure-carrying interfaces. 
%The pressure is set to $\langle \beta \rangle=1.49\%$, by gradually increasing it from the zero-pressure integrable case to obtain an unoptimized, chaotic starting configuration. 
%The field contains a significant volume of islands and chaos, as can be seen from it's Poincar\'e section shown in figure~\ref{fig:initeq}

We first generate an initial coil set that supports the initial equilibrium. 
The geometry of the coils is parametrized by three Fourier series in their Cartesian $x$, $y$, and $z$-coordinates, with the curve describing the $i$-th coil given by: 
\begin{equation}\label{eq:FourCoil}
\begin{aligned}
  x^i(\tau) &= X^i_{c,0} + \sum_{k=1}^K \left[ X^i_{c,k}\cos(k\tau) + X^i_{s,k}\sin(k\tau)\right], \\
  y^i(\tau) &= Y^i_{c,0} + \sum_{k=1}^K \left[ Y^i_{c,k}\cos(k\tau) + Y^i_{s,k}\sin(k\tau)\right], \\
  z^i(\tau) &= Z^i_{c,0} + \sum_{k=1}^K \left[ Z^i_{c,k}\cos(k\tau) + Z^i_{s,k}\sin(k\tau)\right].
\end{aligned}
\end{equation}
We generate a total of five unique coils with Fourier order $K=6$, that are copied according to stellarator symmetry and field periodicity to form a total of 50 coils. 
This results in 195 Fourier modes needed to represent a coil set. 

The coils are initialized as circles enclosing the computational boundary, and the coils are then optimized such that $(\BB_{\rm coils}\cdot\hat{\nn})$ matches the value of $(\BB_{\rm ext}\cdot\hat{\nn})$ from the initial equilibrium on the computational boundary using a FOCUS-like algorithm~\cite{zhu2018designing}. 
This algorithm consists of the gradient-based minimization of the quadratic flux, regularized by a penalty on total coil length, i.e.\ the minimization of: 
\begin{equation}\label{eq:focuslike}
  f_{\rm FOCUS} =  \int_\mathcal{S} \left(\left(\BB_{\rm coils} - \BB_{\rm ext}\right)\cdot \nn \right)^2 \mathrm{d}s  + \mathcal{L}^2_{>L_0}
\end{equation}
where the integral calculates the total quadratic flux through the surface $\mathcal{S}$(= computational boundary) of the difference between a target field $\BB_{\rm ext}$ (given by the SPEC solution) and $\BB_{\rm coils}$ produced by the coils. $\mathcal{L}_{>L_0}$ is a one-sided penalty function that penalizes if the total coil length exceeds $L_0$, and that is zero below $L_0$, with $L_0=1000$m.

An optimum is achieved with a r.m.s. difference between the coil field and the target field of $\delta B/B_0 \simeq 10^{-7}$ with $B_0$ the magnetic field strength on the magnetic axis. 
This produces a SPEC equilibrium indistinguishable from the starting equilibrium in~\cite{baillod2022stellarator}.
In fact figure~\ref{fig:initeq} (a) shows the equilibrium calculated from the coils. 
These initial coils, as well as the normal field they produce on the boundary are shown in figure~\ref{fig:initeq} (b).
We note that the initial field was constructed such that the external field on the computational boundary is exactly opposite to the plasma field (and thus the computational boundary coincides with a magnetic surface) and hence figure~\ref{fig:initeq} (b) shows a non-zero amplitude of the coil field.

\section{Optimizing the equilibrium}\label{sec:brute}
We define our optimization problem with the goal of removing the islands labeled in figure~\ref{fig:initeq}. 
For this we use Cary and Hanson's method~\cite{cary1986stochasticity} of reducing the Greene's Residue~\cite{greene1979method, greene1968two}. 
Greene's residue, $\RR_{m,n}$, of an $(m,n)$ island chain (with $m$ the poloidal and $n$ the toroidal mode number) is a quantity that is calculated at the x- or o-point of said island chain, by calculating the linearized mapping $\mathsf{M}$ of the field line through the x- or o-point. 
The Residue is defined as $\RR_{m,n} = \tfrac{1}{2} - \tfrac{1}{4}\textrm{Tr}(\mathsf{M})$. For details see for example~\citet{greene1968two, smiet2019mapping}.
Greene's residue is larger than zero for o-points, and smaller than zero for x-points, and when the Residue is zero, the $(m,n)$ island chain has reduced to the unbroken $m/n$ magnetic surface. 
The x- and o-points are fixed points of the Poincar\'e map applied $m$ times, and these are found using a Newton method. 
We target the residues of the $\RR_{9/5}$, $\RR_{8/5}$ and $\RR_{7/5}$ island chains, that corresponds to the labeled islands in figure~\ref{fig:initeq}. 

We perform the optimizations using the simsopt stellarator optimization code~\cite{landreman2021simsopt} that has recently been coupled to SPEC. 
We define our degrees-of-freedom for the optimization problem in two different ways: A \emph{brute-force} optimization using all coil degrees-of-freedom, and a reduced-dimensional optimization that we describe in section~\ref{sec:reduced}.
For the \emph{brute-force} optimization, the degrees-of-freedom are the 195 Fourier coefficients that describe the coils according to equation~\ref{eq:FourCoil}.  
The optimization target is a least-squares functional with the following terms: 
\begin{multline}\label{eq:brutesquares}
  f = \mathcal{R}_{9/5}^2 + \mathcal{R}_{8/5}^2 +\mathcal{R}_{7/5}^2 + \\ 0.1(\mathcal{L} - L_0)^2 + 0.1(\mathcal{C}-0.1\mathcal{C}_0^2)
\end{multline} 
where $\mathcal{L}$ is the total coil length, $L_0 = 1010$m, a little larger than the starting coils, and $\mathcal{C}$ is the sum of the mean squared curvatures of the coils for which $\mathcal{C}_0$ is its initial value.

The equilibrium is optimized using a trust-region-reflective least-squares optimization with the Jacobian calculated by a one-directional finite-difference evaluation of the components of equation~\ref{eq:brutesquares}.
The step size for the finite-difference evaluation is set to $10^{-4}$, and the trust region is decreased for degrees-of-freedom corresponding with higher Fourier orders $k$ according to the relation $d_{trust} = 1/20 e^{0.3k}$. 

The last two terms of equation~\ref{eq:brutesquares} were found by trial-and-error, and were found to be necessary for the optimization to converge and to avoid non-convergence of the equilibrium solver when providing unsolvable fields. 
Inspection of the steps taken in optimizations without these terms shows that the optimizer attempts unfeasible configurations, with wild coils caused by high amplitude high order Fourier modes. 
Since both coil length and coil complexity are strongly increased by the higher order Fourier modes, their inclusion in the functional helps taming the steps attempted by the optimizer.
The order-dependent trust region has a similar effect, and with these measures in place, the optimization concludes successfully.

\begin{figure}
  \begin{overpic}[width=\linewidth, percent]{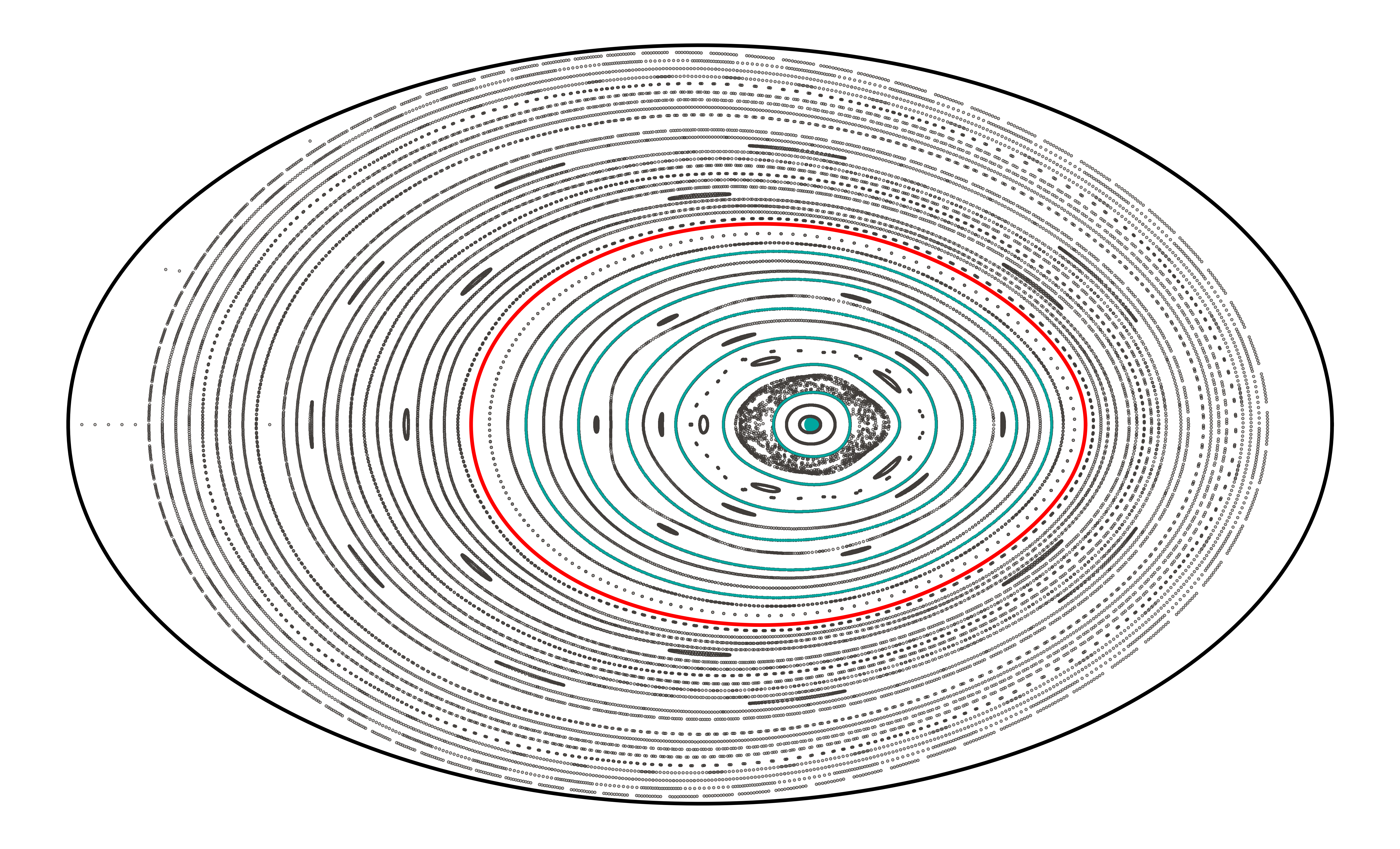}
    \put(5,55){ (a)}
  \end{overpic}
  \begin{overpic}[width=\linewidth, percent]{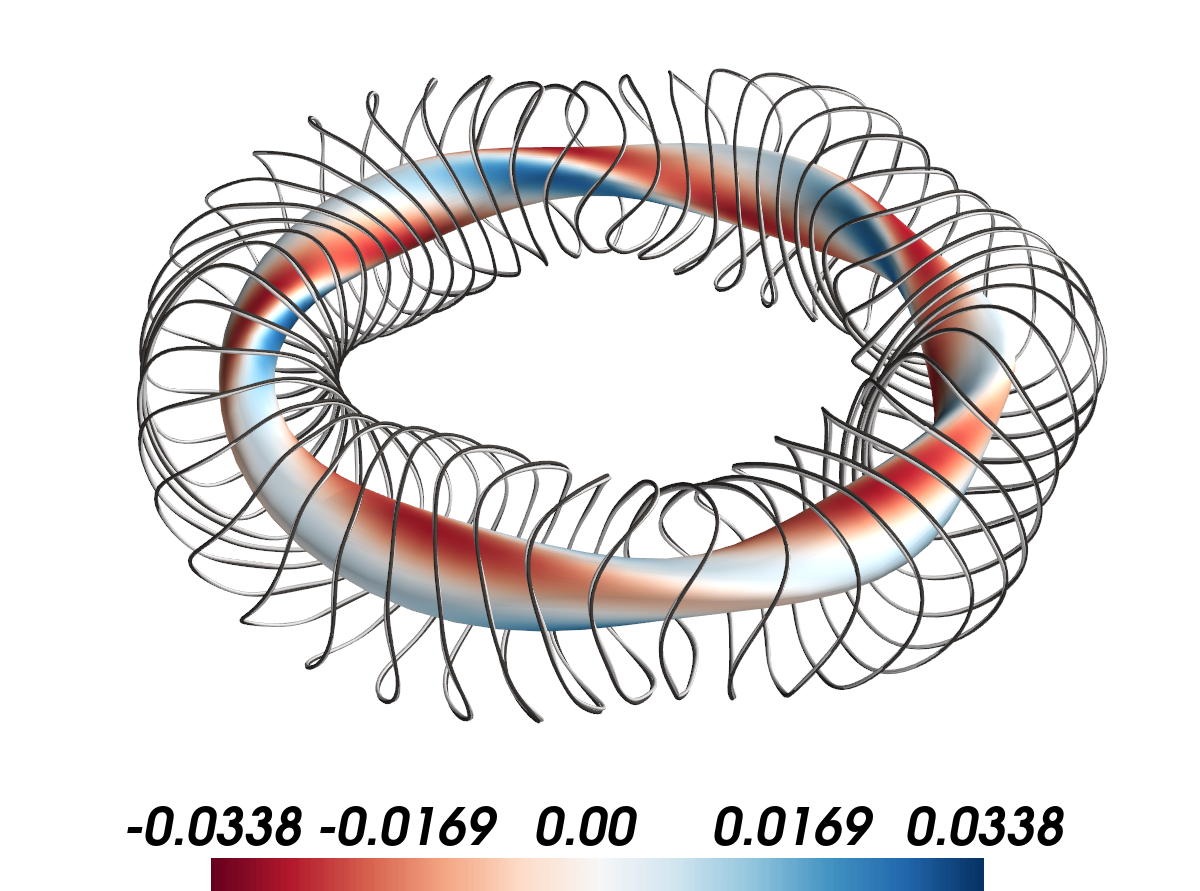}
    \put(5,55){ (b)};
    \put(40,9){\large$(\BB_{\rm coils}\cdot\hat{\mathbf{n}})/B_0$}
  \end{overpic}
  \caption{Optimized equilibrium and coils using the \emph{brute-force} optimization scheme where the degrees-of-freedom are all elements of the coil parametrization. a) Poincar\'e of the equilibrium achieved after optimization. b) The coils that achieve this optimum (black) and the initial coils (white), plotted around the computational boundary (blue-red) colored by the magnitude of the coil field field normal to the surface. The color scale is identical to figures~\ref{fig:initeq},\ref{fig:reducedopt} and~\ref{fig:twostage}. }\label{fig:bruteopt} 
\end{figure}

The optimization achieves an improved coil configuration in which the targeted islands are greatly reduced in size. 
The Poincar\'e section of this field (figure~\ref{fig:bruteopt} (a) ) shows very thin remnants of the islands. 
The optimized coils (\ref{fig:bruteopt} (b), black curves) are only shifted by a small amount with respect to the starting coils (\ref{fig:bruteopt}~(b) and~\ref{fig:initeq}~(b) white curves). 
Figure~\ref{fig:bruteopt} (b) also shows the computational boundary, colored by the normal component of the coil magnetic field on it, using the same color scale as figure~\ref{fig:initeq}.
The optimized coils produce a stronger field on the computational boundary than the initial coils, consistent with the observation that the computational boundary no longer coincides with a magnetic surface. 
We also note that the optimization has starkly reduced the magnetic stochasticity in the Poincar\'e plot, not only in and around the targeted surfaces, but also in the core of the equilibrium. 

The problems encountered in this \emph{brute-force} approach however hint at a fundamental issue with single-stage optimization: 
The optimization problem has 195 degrees-of-freedom.
The free-boundary equilibrium, however, depends only on the field on the computational boundary, which is represented by equation~\ref{eq:FourierDef} with finite Fourier resolution of 110 unique modes. 
This implies that the optimization has an inherent null space, as there will be many linear combinations of coil degrees-of-freedom that cannot be represented by the limited Fourier resolution on the boundary. 
Not only the limited resolution is at issue here; it is a known issue that the coil optimization problem is ill-posed; there are variations of the coil geometry that do not affect the field on the boundary~\cite{landreman2017improved}.
This null space hinders the optimization because it allows for the unfeasible steps taken by the optimizer: variations that do not affect the equilibrium can be freely added to the step.

\section{Reducing the dimensionality of the optimization space}\label{sec:reduced}
In order to mitigate the issues encountered in the \emph{brute-force} optimization, we apply a transformation to the basis of the optimization space. 
The mapping from the coil degrees-of-freedom to the magnetic field on the computational boundary is a mapping from an $I=195$ to a $J=110$-dimensional space: 
\begin{equation}
  g: \mathbb{R}^I \rightarrow \mathbb{R}^J
\end{equation}
$g$ can be linearized around the starting coil set defined by the state vector $\xxc\in\mathbb{R}^I$: 
\begin{equation}
  g(\xxc+\delta\xx) \simeq g(\xxc) + \JJ\delta\xx
\end{equation}
where the Jacobian of the mapping $\JJ_{ij} = \partial g^{(j)} / \partial \xxc^{(i)}$ (note that $(i)$ and $(j)$ are dimension labels) is calculated using one-sided finite-differencing on the evaluation of $g$ with step size $10^{-7}$.

The Jacobian $\JJ$ is a non-square matrix that encodes the effect of coil displacements on the field on the computational boundary.  
We now perform singular value decomposition (SVD) on this Jacobian matrix:
\begin{equation}
  \JJ = \UU \SG \VV.
\end{equation}
SVD decomposes $\JJ$ into two square, real, and orthogonal matrices $\UU$ ($J\times J$) and $\VV$ ($I\times I$) and a diagonal matrix $\SG$ containing the singular values. 
This decomposition has profound significance to the physical problem at hand; rows of $\VV$ form an orthonormal basis of `coil space', and the $i$-th row of $\VV$ (right-singular vector of $\JJ$) is a linear combination of Fourier modes (a displacement of the coils) that effects a change in the normal field given by the $i$-th column of $\UU$, with a transfer efficiency given by the singular value $\SG_{ii}$. 
The SVD algorithm conveniently ranks the singular values from high to low, with the highest singular-value corresponding to the right-singular vector embodying the coil displacement that most efficiently effects change to the field on the computational boundary. 
Because $I>J$, there are many right-singular vectors that have zero singular value; i.e. these are the displacements that do not influence the field on the computational boundary, the null space of our optimization problem! 

\citet{zhu2018hessian} used similar methods in analyzing coil sets at an optimum; they used the singular values of a Hessian matrix to quantify coil sensitivity and identify critical tolerances in construction~\cite{zhu2018hessian,zhu2019identification}. 

We now redefine our optimization problem to use the non-zero right-singular vectors as degrees-of-freedom. 
Further reduction by limiting to a subset of the right-singular vectors could further speed up the computation and will be considered in future work. 
Using all right-singular vectors we have reduced the dimensionality of the problem back to the same dimensionality as needed to represent a doubly-periodic function in our chosen Fourier resolution, i.e. 110 degrees-of-freedom. 
This is the minimal dimensionality for a first stage optimization in a two-stage process, as it is the minimal dimensionality of a  fixed-boundary representation~\cite{henneberg2021representing} with similar spectral complexity.
Our reduced representation furthermore removes coil displacements that do not affect the field on the boundary, i.e. it removes the inherent null space in coil optimization~\cite{landreman2017improved}.

In this new optimization space we no longer need the regularizing penalties on total coil length and coil complexity, and we optimize the stellarator equilibrium with the least-squares functional: 
\begin{equation}\label{eq:reducedfunctional}
  f = \mathcal{R}_{9/5}^2 + \mathcal{R}_{8/5}^2 +\mathcal{R}_{7/5}^2,
\end{equation}
without the terms on the coils present in equation~\eqref{eq:brutesquares}.

\begin{figure}
  \begin{overpic}[width=\linewidth, percent]{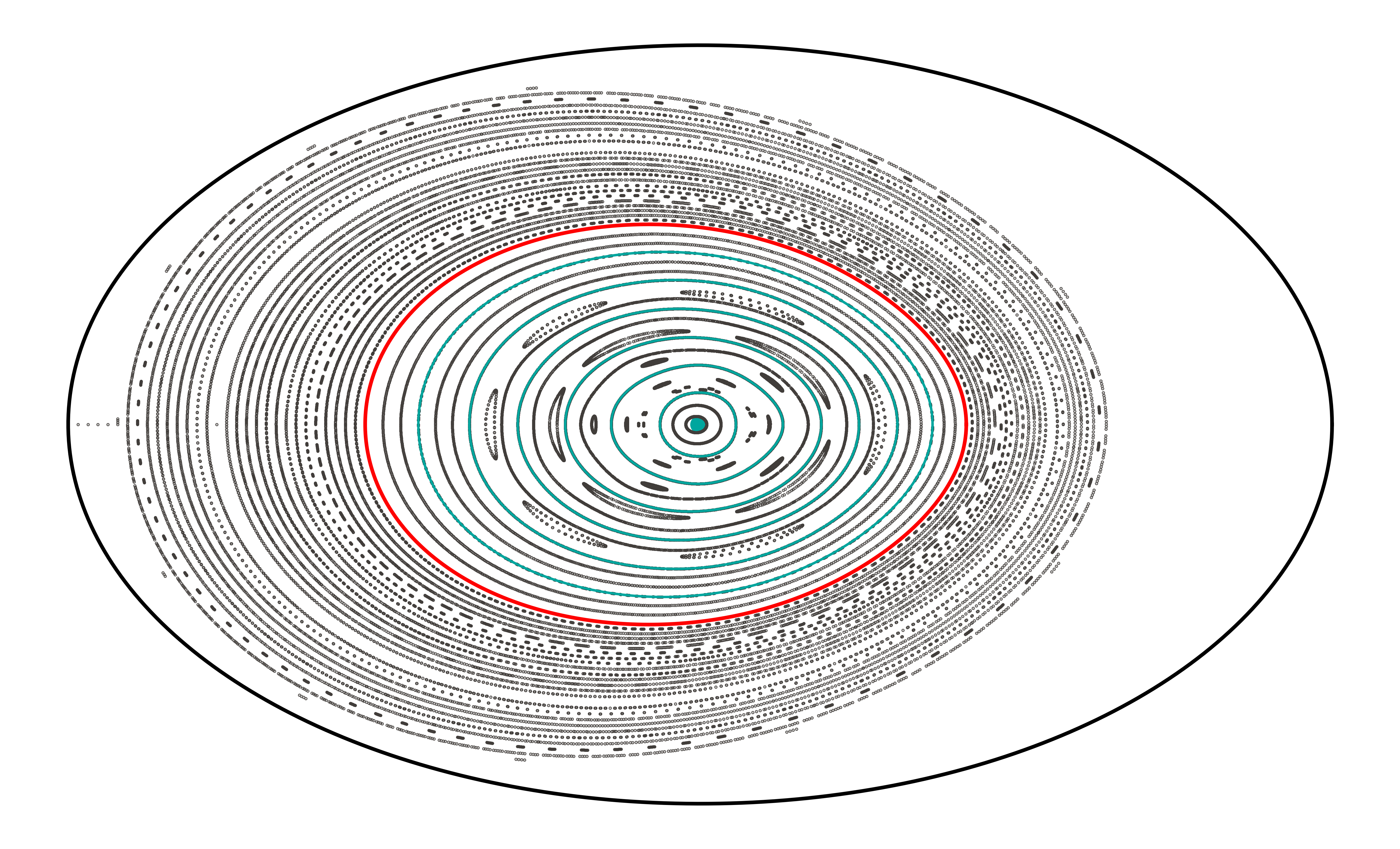}
    \put(5,55){ (a)}
  \end{overpic}
  \begin{overpic}[width=\linewidth, percent]{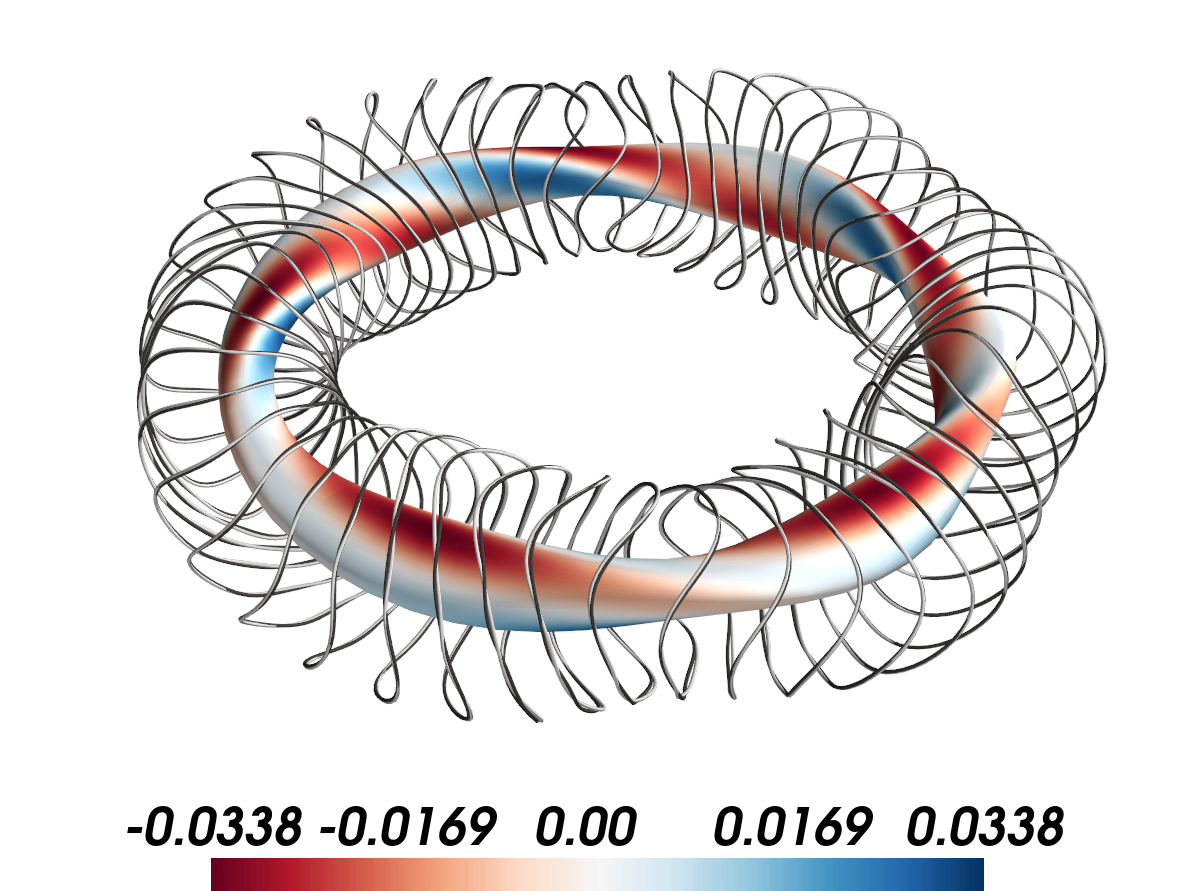}
    \put(5,55){ (b)}
    \put(40,9){\large$(\BB_{\rm coils}\cdot\hat{\mathbf{n}})/B_0$}
  \end{overpic}
  \caption{Optimized equilibrium and coils using the dimensionality-reduced optimization scheme. a: Poincar\'e section of the field. b: The coils that achieve this optimum (black) and the initial coils (white), plotted around the computational boundary (blue-red) colored by the magnitude of the field normal to the surface. The color scale is identical to figures~\ref{fig:initeq},\ref{fig:bruteopt} and~\ref{fig:twostage}.}\label{fig:reducedopt}
\end{figure}

In the optimized equilibrium, the targeted island chains can no longer be seen at all, as shown in figure~\ref{fig:reducedopt} (a). 
The entire equilibrium is also shifted from its original position, and the computational boundary is clearly no longer a flux surface. 
The coils are not significantly displaced from their starting positions, as shown in figure~\ref{fig:reducedopt} (b). 
The normal component of the coil field on the computational boundary, shown by the color in figure~\ref{fig:reducedopt} (b) is increased over the \emph{brute-force} optimization. 
The achieved optimum is an order of magnitude more performant than the \emph{brute-force} approach, as can be seen from the optimization results in table~\ref{tab:opresults}. 
From this table we can also see that the coil complexity is increased by less than 2\% and the total coil length is even reduced, though these were not optimization targets.  

The reduced-dimensional optimization space has significantly increased the robustness of the calculation. 
The \emph{brute-force} optimization functional in equation~\eqref{eq:brutesquares} was found through trial-and-error. 
Optimization functionals without the stabilizing weights on coil complexity and length would not converge, get stuck in local minima, or lead to crashes of the equilibrium solver. 
The reduced-dimensional optimization has no need for stabilizing terms. 

\section{Comparison with two-stage optimization}
\begin{figure*}
  \centering
  \begin{minipage}{.55\textwidth}
    \centering
    \begin{overpic}[width=\textwidth, percent]{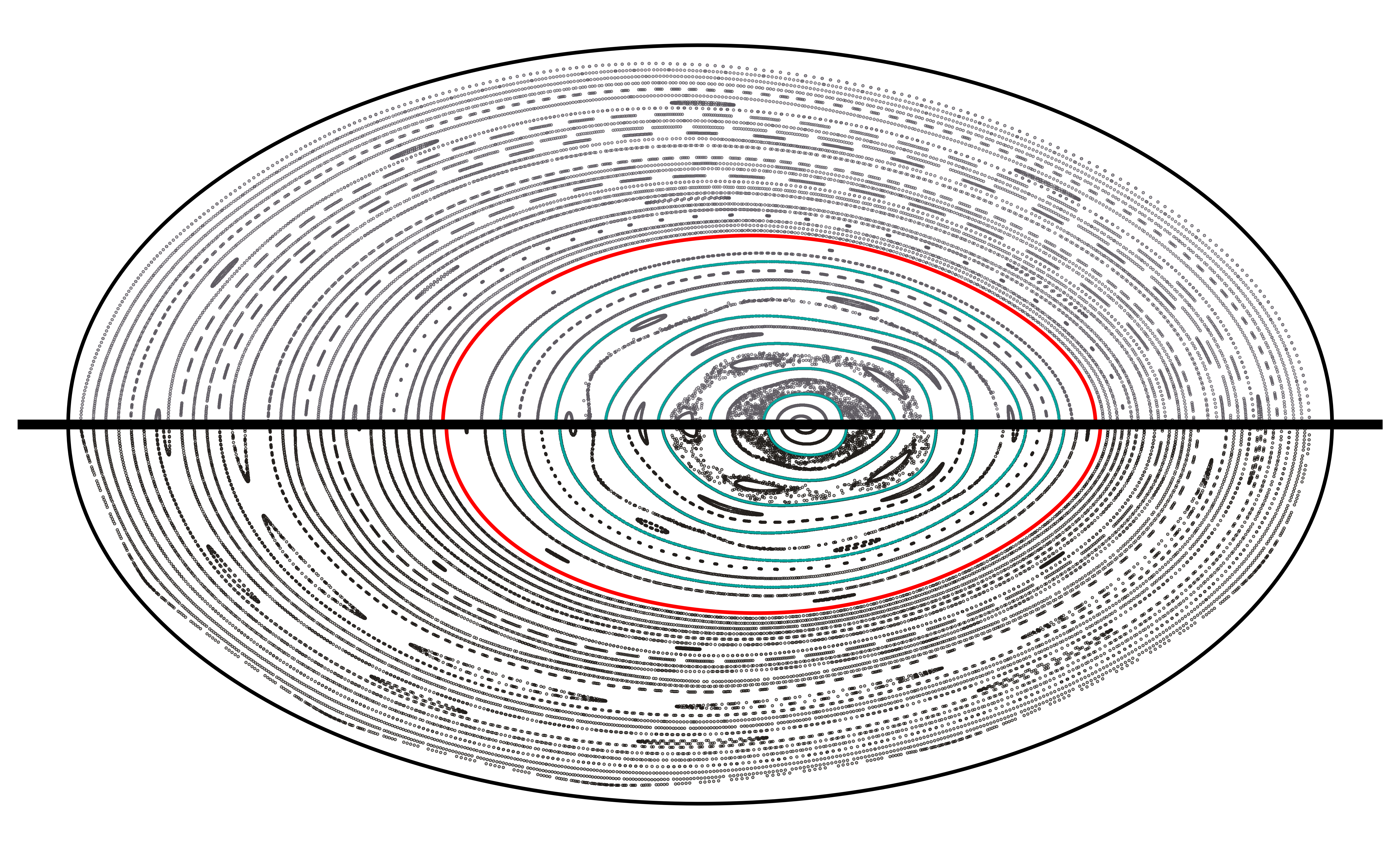}
    \put(5,55){ (a)}
    \put(5,52){ first stage}
    \put(5,5){ second stage}
    \end{overpic}
    \begin{overpic}[width=\textwidth]{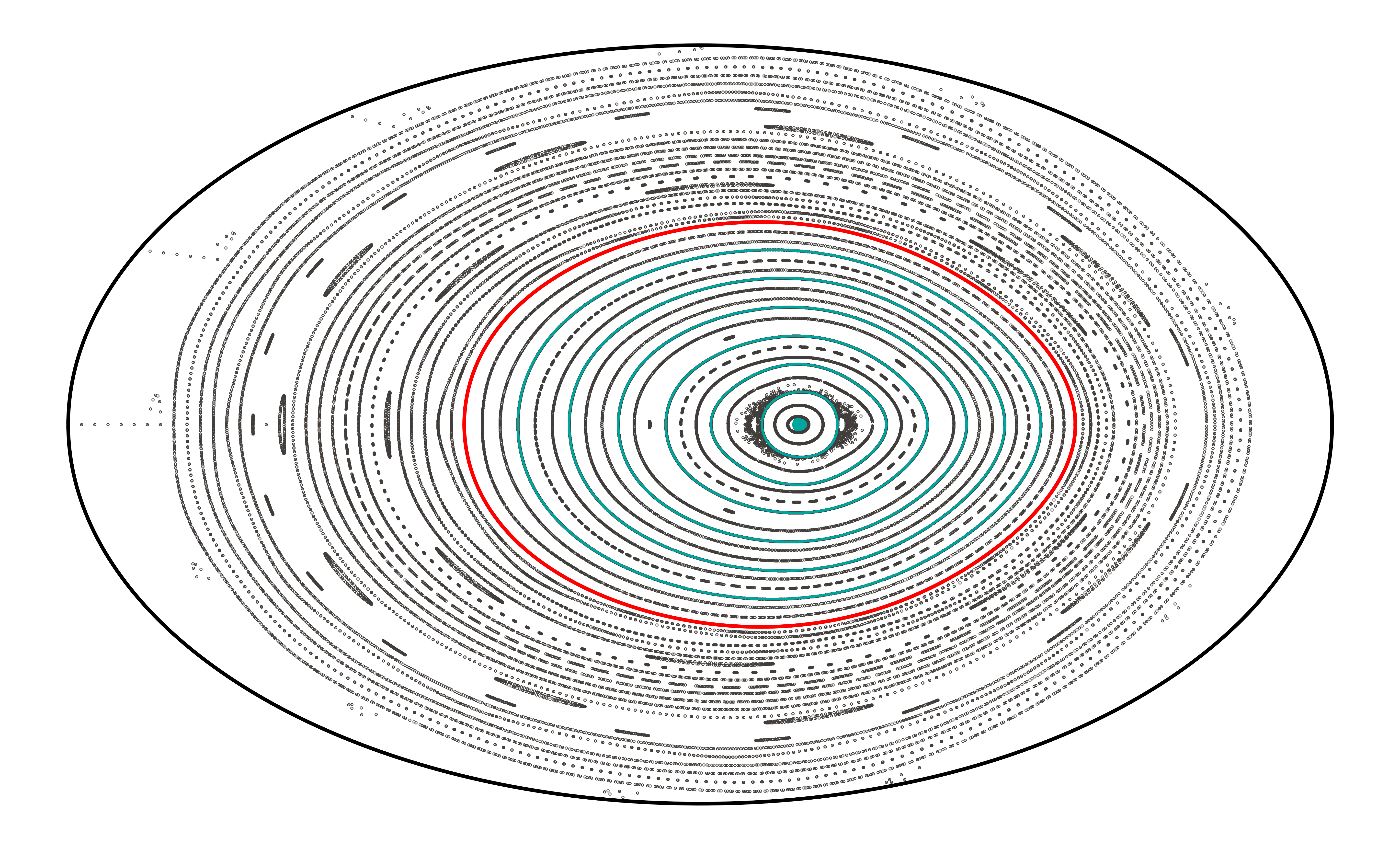}
    \put(5,55){ (c)}
    \put(5,5){single-stage}
    \end{overpic}
  \end{minipage}
  \hfill
  \centering
  \begin{minipage}{.44\textwidth}
    \centering
    \begin{overpic}[width=\textwidth]{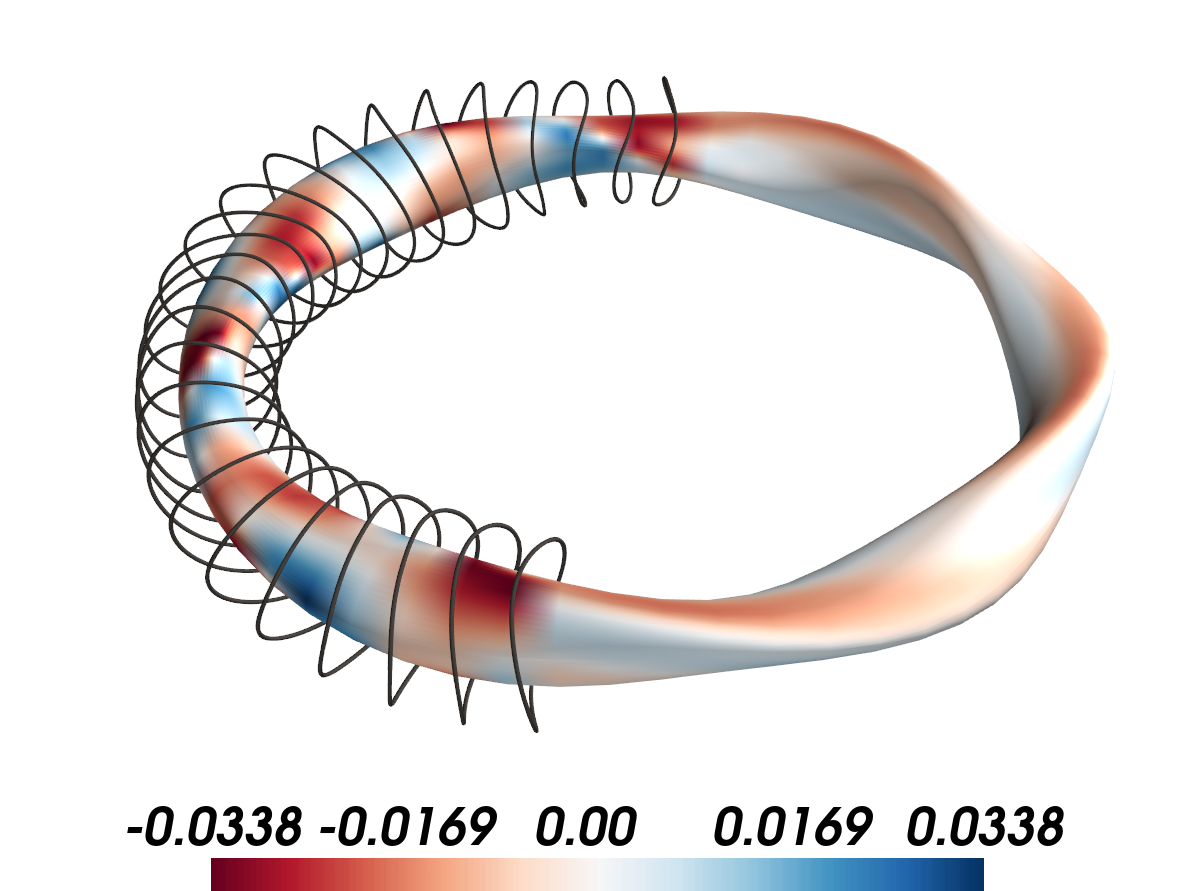}
      \put(5,55){(b)}
      \put(25,40){second stage}
      \put(65,40){target field}
      \put(40,9){\large$(\BB_{\rm coils, ext}\cdot\hat{\mathbf{n}})/B_0$}
    \end{overpic}
    \begin{overpic}[width=\textwidth]{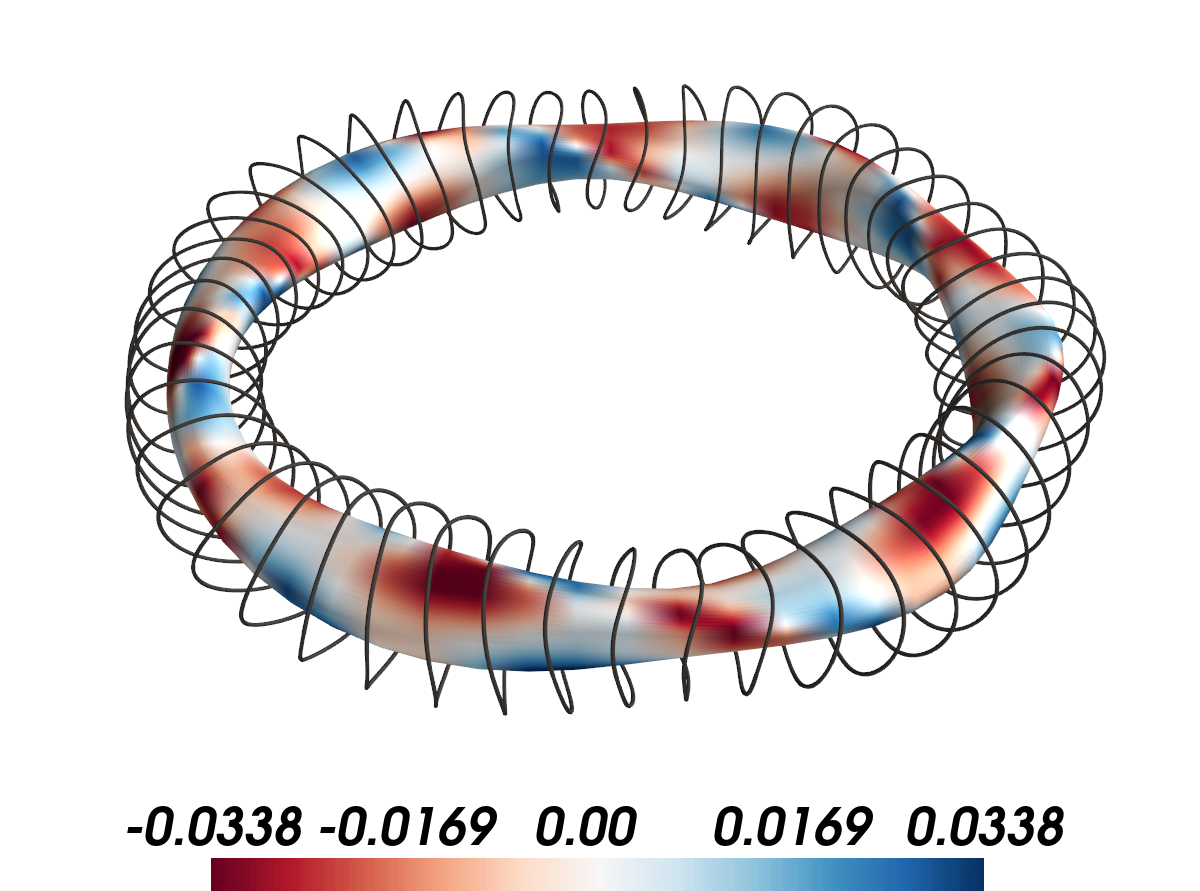}
    \put(5,55){(d)}
    \put(40,9){\large$(\BB_{\rm coils}\cdot\hat{\mathbf{n}})/B_0$}
    \end{overpic}
  \end{minipage}
  \hfill
%  \begin{minipage}{.23\textwidth}
%    \centering
%    \includegraphics[width=\textwidth]{TWOSTAGE_coils.png}
%    \put(0.1\textwidth,0.9\textheight){\tiny (b)}
%    \oppoverlay{
%      \put(0.5\textwidth,0.5\textheight){\textcolor{blue}{text for image b}}
%    }
%  \end{minipage}
%  \hfill
%  \begin{minipage}{.33\textwidth}
%    \centering
%    \includegraphics[width=\textwidth]{SHORTER_poincare.png}
%    \put(0.1\textwidth,0.9\textheight){\tiny (c)}
%    \oppoverlay{
%      \put(0.5\textwidth,0.5\textheight){\textcolor{green}{text for image c}}
%    }
%  \end{minipage}
  \caption{Comparison of single-stage and two-stage optimization of islands in stellarator equilibria. a: Poincar\'e section of the single-stage optimized field (top half a), and of the second-stage field consistent with the coils (bottom half a). The islands that were mostly removed in the first stage are inadvertently re-introduced in the second stage.  
  b) coils of the two-stage optimized field (left side) and the resultant normal field to the computational boundary, and the target thereof from the single-stage optimization (right half).
  c) Poincar\'e section of the single-stage optimized field.
  d: single-stage optimized coils.
  The color scale in c and d is identical to figures~\ref{fig:initeq}-\ref{fig:reducedopt}.}
  \label{fig:twostage}
\end{figure*}
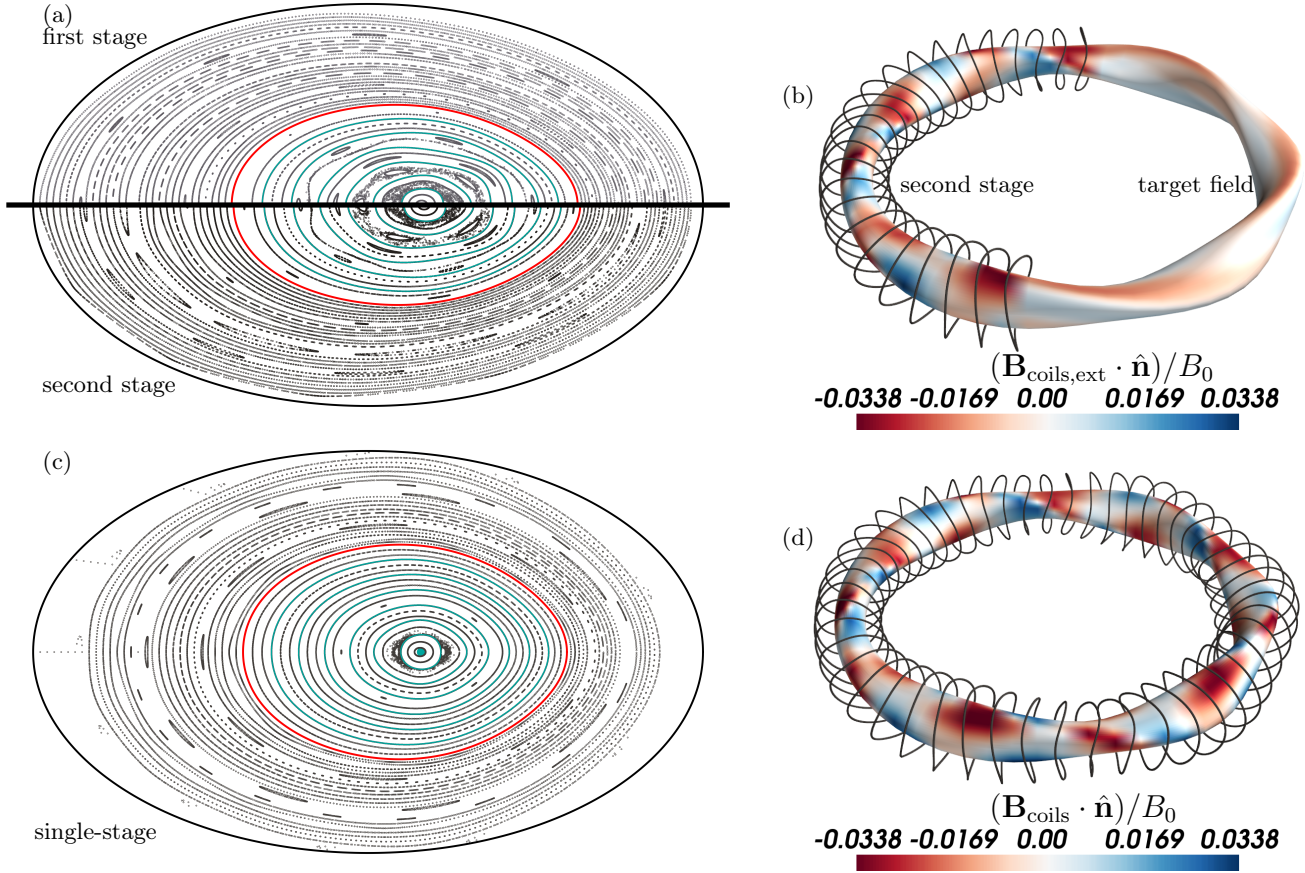

We finally demonstrate the advantage of single-stage optimization by performing a similar optimization in two stages. 
To increase the realism of this example, we generate initial coils that are 40\% shorter (coils closer to the plasma than the coils in section I), with a total length of $600$m.
For the first stage of this optimization, we repeat the island optimization procedure laid out in~\cite{baillod2022stellarator}.
Our degrees-of-freedom are the 110 Fourier components describing the magnetic field normal to the computational boundary, and the least-squares functional is identical to equation~\eqref{eq:reducedfunctional}.
Similar to the results of~\citet{baillod2022stellarator}, this optimization yields an equilibrium in which two new island chains appear with mode numbers $(m,n) = (10,5)$ and $(6,5)$ respectively. 
A second optimization is performed on this intermediate optimum with the least-squares functional 
\begin{equation}
  f = \mathcal{R}_{10/5}^2 + \mathcal{R}_{9/5}^2 + \mathcal{R}_{8/5}^2 +\mathcal{R}_{7/5}^2 +  \mathcal{R}_{6/5}^2.
\end{equation}
The optimization runs successfully and the Poincar\'e section of this field is shown in figure~\ref{fig:twostage} (a) (top half). 
The island size is reduced, but the optimum achieved is not as performant as the single-stage or the reduced single stage optimizations.

\begin{table*}
  \centering
  \caption{Optimization Results\label{tab:opresults}}
  \begin{tabular}{c|ccccc} % ten columns with numbers

Optimization & $\RR_{9/5}$ & $\RR_{8/5}$ & $\RR_{7/5}$ & $\mathcal{L}$ & $\mathcal{C}$ \\ 
\hline
  initial configuration           & 7.7$\times 10^{-2}$     & -2.11$\times 10^{-1}$  & 2.95$\times 10^{-1}$    & 1000  & 0.995 \\ 
  \emph{brute-force} optimization & 3.2$\times 10^{-3}$     & -1.78$\times 10^{-2}$  & 5.4 $\times 10^{-2}$    & 1010  & 0.847 \\ 
 reduced-dimensional optimization & -5.32$\times 10^{-4}$   & 7.78$\times 10^{-3}$   & -2.35$\times 10^{-3}$ & 997.7 & 1.02 \\ 
first stage (shorter coils)       & -4.23$\times 10^{-3}$   & -1.07$\times 10^{-2}$  & 7.77$\times 10^{-3}$  &       &  \\ 
second stage (shorter coils)      & 2.61$\times 10^{-2}$    & -5.81$\times 10^{-2}$  & 5.22$\times 10^{-2}$   & 600   & 1.7 \\ 
single-stage (shorter coils)      & 1.49$\times 10^{-2}$    & 8.97$\times 10^{-3}$   & 9.39$\times 10^{-3}$  & 600.2 & 1.72 \\ 
\end{tabular}

%
% initial configuration            & 0.077     & -0.211  & 0.295    & 1000  & 0.995 \\ 
% brute-force optimization         & 0.0032    & -0.0178 & 0.054    & 1010  & 0.847 \\ 
% reduced-dimensional optimization & -0.000532 & 0.00778 & -0.00235 & 997.7 & 1.02 \\ 
%first stage (shorter coils)       & -0.00423  & -0.0107 & 0.00777  &       &  \\ 
%second stage (shorter coils)      & 0.0261    & -0.0581 & 0.0522   & 600   & 1.7 \\ 
%single-stage (shorter coils)      & 0.0149    & 0.00897 & 0.00939  & 600.2 & 1.72 \\ 
%{\textwidth}{
%  @{\extracolsep{\fill}}% fill the space between columns
%  l % one left aligned column
%  *{10}{S[table-format=-1.2]} % ten columns with numbers
%}
\end{table*}

This first-stage optimized equilibrium is then used as a target for a second-stage coil optimization which seeks to match the field on the computational boundary.
We start from a set of circular coils, and we use the FOCUS-like algorithm described in section~I.  
To prevent the optimizer from getting stuck in a local minimum, the optimization is performed four times; first with $\mathcal{L}_0=600$, then the optimized coils are re-optimized with $\mathcal{L}_0=610$, and then twice more with $\mathcal{L}_0=600$. 

The Poincar\'e section of the equilibrium after the second stage optimization is shown in figure~\ref{fig:twostage} (a) (bottom half). 
We see that the islands that were mostly removed in the first stage optimization are re-introduced in the second stage.
The coils from the second stage are shown in figure~\ref{fig:twostage}~(b) (left half), together with the normal field they produce on the computational boundary. 
The right half of figure~\ref{fig:twostage}~(b) shows the target field, namely the result of the first stage that needs to be matched by the second stage, but a close match is not possible due to unavoidable coil ripple. 

We compare this result to the equivalent single-stage optimization. 
First, coils are generated for the initial equilibrium with a single optimization with $\mathcal{L}_0=600$.
We then use our dimensionality reduction scheme to generate a 110-dimensional optimization space from this initial coil position. 
The target functional for the single-stage optimization consists only of the three initial island chain Residues (i.e. equation~\eqref{eq:reducedfunctional}). 
The optimization is performed similarly to the other reduced optimization, and runs to completion after two Jacobian evaluations. 

The reduced optimization of the shorter coil set is successful, resulting in a reduced island width visible in the Poincar\'e section (figure~\ref{fig:twostage} (c) ).
This single-stage optimization yields better results than the two-stage process; table~\ref{tab:opresults} shows that the residues are roughly halved. 
The coils of the single-stage optimization show significant ripple on the computational boundary, as shown in figure~\ref{fig:twostage} (d), comparable to the ripple of the two stage optimized coils, yet the island residues are significantly decreased. 
We postulate that this is due to the fact that the optimizer can distribute the coil ripple on the computational boundary in a way that has minimal effect on the islands, whereas when the second-stage targets an already optimized field, the mismatch is distributed randomly, leading to a larger effect on the islands. 
We also note that the Poincar\'e section of the single-stage optimized field shows significantly reduced chaos, also in the core of the equilibrium near the axis, hinting that the single-stage process changes the field in a more robust way, and is able to find solutions that the two-stage process cannot.

\section{Conclusions and discussion}
We have demonstrated a first of a kind single-stage optimization of a finite-$\beta$ stellarator equilibrium with islands. 
We have furthermore shown how the dimensionality of a single-stage optimization can be reduced to that of the two-stage problem, and that this reduction scheme improves the optimization performance by removing a null space. 
%It is important to note that even though we have reduced our problem to the same \emph{dimensionality}, it is not reduced it to the same computational \emph{complexity}; The first stage can be performed with fixed-boundary equilibrium solves, which is several times faster than free-boundary equilibrium solves.
%In SPEC the freeboundary solve is efficiently implemented by only requiring the field on the computational boundary compared to VMEC which requires the field in the entire volume, requiring an expensive Biot-Savart calculation. 

All optimizations were performed on a cluster using a total of 288 cores per run, split into 24 groups of 12 cores. 
The brute-force optimization took 4000 CPU-hours to complete, whereas the reduced-dimensional optimization took 3500 cpu-hours to terminate. 
The optimizations are terminated when the step size becomes too small. 
The reduced-dimensional optimization evaluated significantly more finite-difference Jacobians (the computationally most costly step), and as a result reached a significantly better optimum configuration. 
The total wall clock time was on the order of 10 hours.

This dimensionality reduction scheme has broader applicability than the problem at hand, and we expect it to to enhance most single-stage optimizations, as the number of degrees-of-freedom to describe a coil set are generally larger than those needed to describe the plasma boundary. 
It can be further improved in many ways which will be the subject of future research, including but not limited to: 
\begin{enumerate}
  \item Using only a subset of the singular vectors.
  \item Biasing the Jacobian calculation so the most relevant (to the problem at hand) coil displacements have largest singular values.
  \item Projecting the reduced degrees-of-freedom into a subspace that conserves desirable coil metrics such as total length or coil complexity. 
\end{enumerate}

This scheme is furthermore not limited to any representation of the coils. 
The Fourier representation in~\eqref{eq:FourierDef} can be replaced with any other, for example spline-based~\cite{lonigro2022stellarator, takahashi2024designing}, coil representations. 
Although splines have the advantages of having local support, less degeneracy in the representation, and similar magnitude effect on the produced field for each degree-of-freedom, coils represented in this way will still have the inherent null space consisting of displacements that do not affect the field at the plasma~\cite{landreman2017improved, landreman2016efficient}.

The dimensionality reduction scheme is most efficient when one can assume the that the linearization embodied by the Jacobian $\JJ$ remains valid, i.e.\ the coils are only displaced a small amount.
This makes the method very well suited for the problem of removing islands for which only minor adjustments to the coils are sufficient. 
If the coils need to be significantly displaced, or if the coils are optimized starting from a simplified ('cold start') configuration, the linearization on which we performed the SVD can no longer be assumed to be valid. 
In such a case, the mapping could be re-calculated regularly. 
Since the calculation of the Jacobian $\JJ$ (from coil degrees-of-freedom to field-on-boundary) does not involve free-boundary solves and can be evaluated in seconds, this can still significantly speed up an optimization.

The capability to include islands in a single-stage finite-$\beta$ stellarator optimization opens up new possibilities for stellarator design. 
We demonstrated how this scheme can be used to remove unwanted islands, but specific islands can now also be intentionally introduced. 
One example is the optimization of either an island divertor or a nonresonant divertor~\cite{bader2017hsx}. 
The capability to resolve the magnetic structure at finite $\beta$ directly from from the coils that support it, will make it possible to enforce the necessary space for a divertor solution, and ensure its proper functioning in future reactors.

We have demonstrated our single-stage optimization in a simplified geometry, consisting of a rotating ellipse. 
In future work we will apply our techniques to more strongly shaped and reactor-relevant stellarator geometries

\begin{acknowledgments}
This work was supported by a grant from the Simons Foundation\ (1013657, JL). 
This work has been carried out within the framework of the EUROfusion Consortium, via the Euratom Research and Training Programme (Grant Agreement No 101052200 — EUROfusion) and funded by the Swiss State Secretariat for Education, Research and Innovation (SERI). Views and opinions expressed are however those of the author(s) only and do not necessarily reflect those of the European Union, the European Commission, or SERI. Neither the European Union nor the European Commission nor SERI can be held responsible for them. 
\end{acknowledgments}

\bibliography{references}% Produces the bibliography via BibTeX.

\end{document}